\documentclass[twocolumn, 10pt, pre, aps, showpacs, reprint, amsmath, amssymb, superscriptaddress]{revtex4-1}

\usepackage[pdftex]{graphicx} 
\usepackage{bm}
\usepackage{subfigure}

\begin{document}
\newcommand{\kt}{\gamma}
\newcommand{\lzt}{q_z}
\newcommand{\atled}{\bm{\nabla}}
\newcommand{\dx}{\frac{\partial}{\partial_x}}
\newcommand{\dy}{\frac{\partial}{\partial_y}}
\newcommand{\dz}{\frac{\partial}{\partial_z}}
\newcommand{\dt}{\frac{\partial}{\partial_t}}
\newcommand{\sqrdt}{\frac{\partial^2}{\partial_t^2}}
\newcommand{\pbyp}[2]{\frac{\partial #1}{\partial #2}}
\newcommand{\dbyd}[2]{\frac{d #1}{d #2}}
\newcommand{\ex}{\bm{e}_x}
\newcommand{\ey}{\bm{e}_y}
\newcommand{\ez}{\bm{e}_z}
\newcommand{\besselj}[2]{\mathrm{J}_{#1}(#2)}
\newcommand{\besseljp}[2]{\mathrm{J'}_{#1}(#2)}
\newcommand{\hankel}[3]{\mathrm{H}_{#1}^{(#2)}(#3)}
\newcommand{\hankelp}[3]{\mathrm{H'}_{#1}^{(#2)}(#3)}
\newcommand{\laplace}{\Delta}
\newcommand{\neff}{n_{\mathrm{eff}}}
\newcommand{\fexp}{f_{\mathrm{expt}}}
\newcommand{\ftheo}{f_{\mathrm{calc}}}
\newcommand{\nexp}{\tilde{n}}
\newcommand{\Gtheo}{\Gamma_{\mathrm{calc}}}
\newcommand{\Gexp}{\Gamma_{\mathrm{expt}}}
\newcommand{\Grad}{\Gamma_{\mathrm{rad}}}
\newcommand{\Gabs}{\Gamma_{\mathrm{abs}}}
\newcommand{\Gant}{\Gamma_{\mathrm{ant}}}
\newcommand{\reffig}[1]{\mbox{Fig.\ \ref{#1}}}
\newcommand{\refsec}[1]{\mbox{Sec.\ \ref{#1}}}
\newcommand{\subreffig}[1]{\mbox{Fig.\ \subref{#1}}}
\newcommand{\refeq}[1]{\mbox{Eq.\ (\ref{#1})}}
\renewcommand{\Re}[1]{\mathrm{Re}(#1)}
\renewcommand{\Im}[1]{\mathrm{Im}(#1)}

\newcommand{\EzWP}{E_z^\mathrm{(WP)}}
\newcommand{\slit}{s}
\newcommand{\dist}{d}

\hyphenation{de-ve-lop-ment si-mu-la-tion cor-res-pon-ding cor-res-pon-ding-ly nu-me-ri-cal lea-ving dy-na-mic dy-na-mics lea-kage sho-wing dif-fe-ren-ce pa-nel pa-nels vi-ci-ni-ty pe-rio-dic im-me-dia-te-ly ini-tial mo-ving in-ter-fe-ren-ce pro-pa-ga-tor eva-lua-ted re-sul-ting pro-pa-ga-tion ma-ni-fest-ly per-pen-di-cu-lar rec-tan-gu-lar re-so-nance an-echoic re-gu-lar re-so-na-tors va-nish ana-lo-gy con-fi-gu-ra-tion}

\title{Double-Slit Experiments with Microwave Billiards}

\author{S. Bittner}
\author{B. Dietz}
\email{dietz@ikp.tu-darmstadt.de}
\author{M. Miski-Oglu}
\author{P. Oria Iriarte}
\affiliation{Institut f\"ur Kernphysik, Technische Universit\"at Darmstadt, D-64289 Darmstadt, Germany}
\author{A. Richter}
\email{richter@ikp.tu-darmstadt.de}
\affiliation{Institut f\"ur Kernphysik, Technische Universit\"at Darmstadt, D-64289 Darmstadt, Germany}
\affiliation{ECT*, Villa Tambosi, I-38123 Villazzano (Trento), Italy}
\author{F. Sch\"afer}
\affiliation{LENS, University of Florence, I-50019 Sesto-Fiorentino (Firenze), Italy}
\date{\today}

\begin{abstract}
Single and double-slit experiments are performed with two microwave billiards with the shapes of a rectangle and a quarter stadium, respectively. The classical dynamics of the former is regular, whereas that of the latter is chaotic. Microwaves can leave the billiards via slits in the boundary, forming interference patterns on a screen. The aim is to determine the effect of the billiard dynamics on their structure. For this the development of a method for the construction of a directed wave packet by means of an array of multiple antennas was crucial. The interference patterns show a sensitive dependence not only on the billiard dynamics but also on the initial position and direction of the wave packet.
\end{abstract}


\maketitle

\section{\label{introduction}Introduction}

At the beginning of the 19th century, Thomas Young performed for the first time an experiment that still attracts strong interest in physics: the interference of light beams passing through a double slit \cite{Young1845}. This experiment and its outcome have played a profound role in the development of optics and quantum mechanics and have been used since as a paradigm to unveil the wave nature of a number of physical entities, in particular single electrons \cite{Jonsson1961, Tonomura1989}, neutrons \cite{Zeilinger1988}, atoms \cite{Carnal1991, Noel1995} and molecules  \cite{Arndt1999}.

Recently, Casati and Prosen carried out a numerical simulation of a double-slit experiment with a well directed Gaussian wave packet initially confined inside quantum billiards \cite{Casati2005}, whose classical dynamics are regular or chaotic. The time evolution of the wave packet is governed by the first-order time-dependent Schr{\"o}dinger equation. The wave packets may leak out of the billiard via two slits in its sidewall. The interference patterns resulting from diffraction of the wave packet at the slit openings have been investigated. For the billiard with regular dynamics, the interference pattern is similar to the well-known result obtained from a double-slit experiment with plane waves in the Fraunhofer regime (far field region); i.e., the intensity encountered on a screen is equal to the sum of the intensities from the two corresponding single-slit experiments plus an interference term \cite{Hecht2002}. On the other hand, for the billiard with chaotic dynamics, the intensity on the screen is equal to the superposition of the diffraction maxima radiated from each of the single slits. As a consequence, the intensity pattern becomes unimodal. 

The experiments described in the present work were initially motivated by these numerical studies. They were performed with flat cylindrical microwave cavities (also called microwave billiards) that have one or two openings in their side walls.  Below a certain excitation frequency, only the lowest transverse magnetic modes with electric field perpendicular to the top and bottom plate of the cavity exist. Accordingly, there the related Helmholtz equation is identical with the time-independent Schr{\"o}dinger equation of the quantum billiard of corresponding shape \cite{Richter1999, StoeckmannBuch2000}. An appropriate choice of the shape of the cavity allows the investigation of wave phenomena in quantum billiards with regular, chaotic, or mixed dynamics. In contrast to the quantum case, the electromagnetic wave equation governing the time evolution of the waves in microwave billiards is of second order. Still, as will be outlined below, the experiments provide insight into the dependence of the interference patterns resulting from such double-slit experiments on the classical dynamics. 

The paper is organized as follows. Section \ref{sec:exp} is devoted to the description of the experimental setup. Stationary wave patterns originating from the diffraction of electromagnetic waves emitted from a single antenna, which is situated inside the microwave billiard, at the openings are presented in \refsec{stationary}. In \refsec{temp} we investigate the temporal evolution of the diffracted waves in the exterior close to the slits. In \refsec{gaussian} we present a new method developed for the construction of a directed initial wave packet with an array of emitting antennas. It is used to realize the situation considered in the numerical simulation of Ref.\ \cite{Casati2005}.

\section{\label{sec:exp}Experiments}
\subsection{\label{ssec:ExpBill}Billiards}

\begin{figure}[tb]
\begin{center}
\subfigure[]{
	\includegraphics[width = 8 cm]{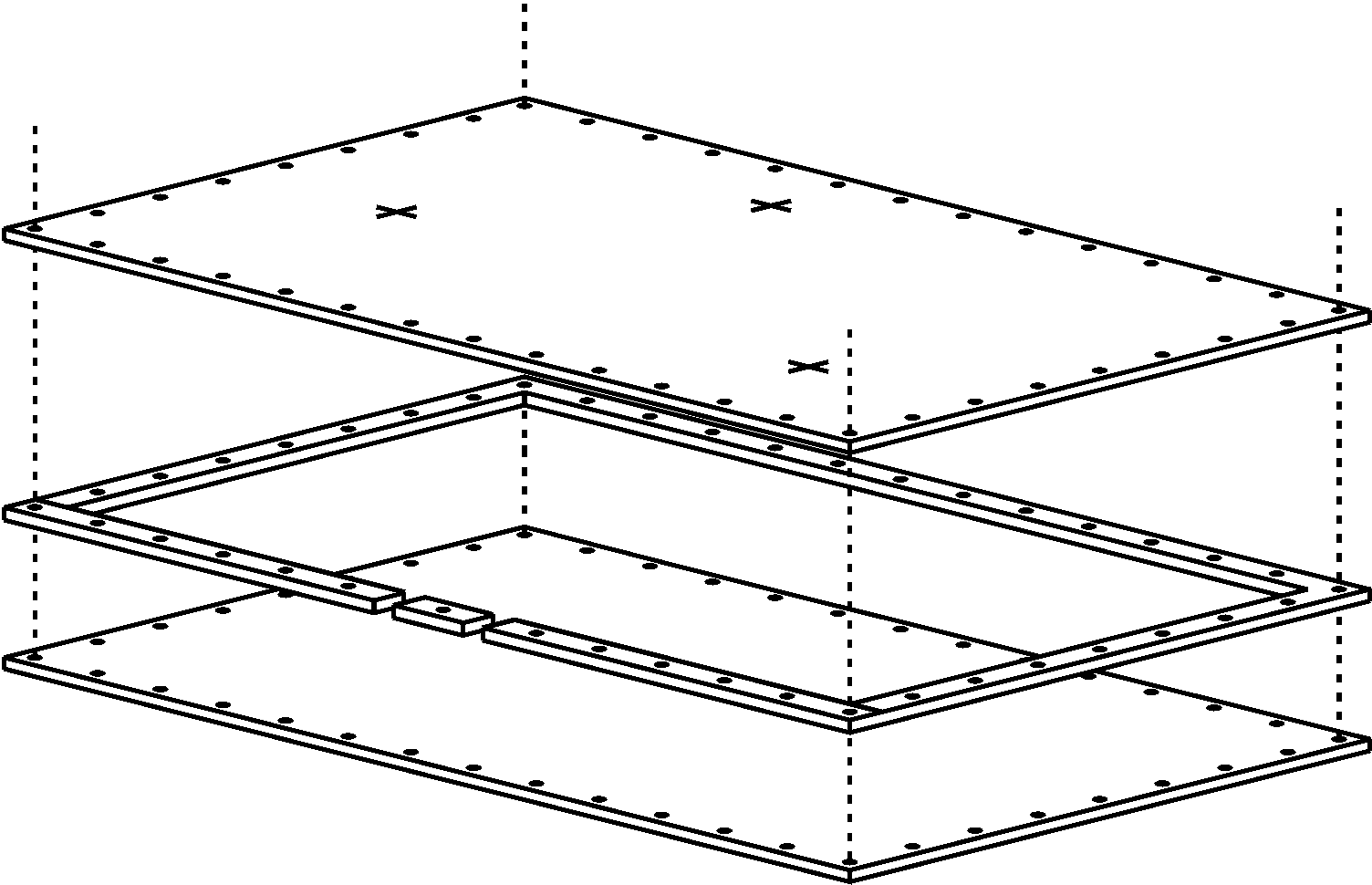}
	\label{sfig:ExpSetup3d}
}
\subfigure[]{
	\includegraphics[width = 8 cm]{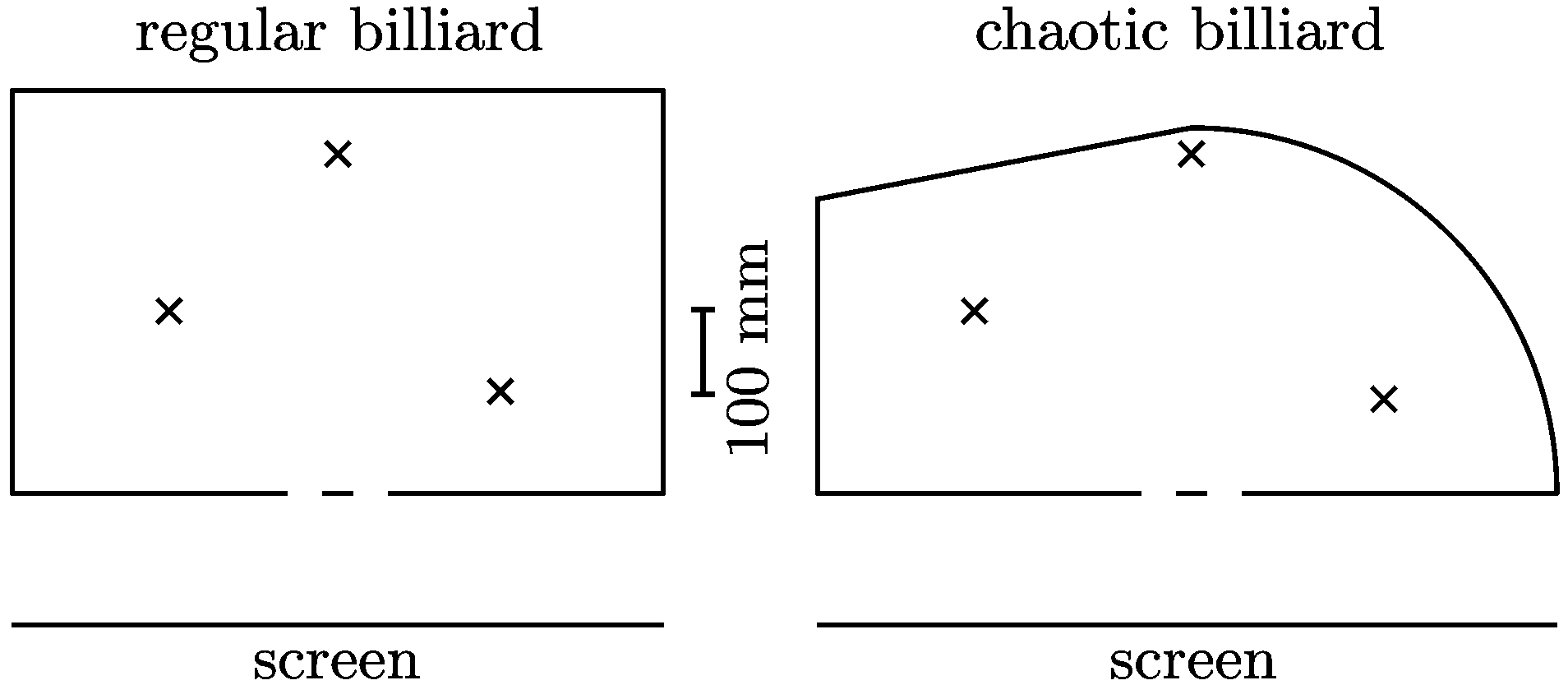}
	\label{sfig:ExpSetup2d}
}
\caption{\subref{sfig:ExpSetup3d} Perspective depiction of the modular assembly of a rectangular microwave billiard (not to scale). The microwave resonator is composed of a frame, $5$ mm high, defining the boundary of the billiard squeezed between a top and bottom plate, each $5$ mm thick. These are screwed together via the holes along the boundaries. Small holes in the top plate (indicated by the crosses) are used to introduce the antennas. \subref{sfig:ExpSetup2d} Sketch of the experimental setup. The interference pattern of the waves emanating from a billiard with two slits are detected on a ``screen.'' A rectangular billiard with regular dynamics and a tilted stadium billiard with chaotic dynamics are used in the experiments. The positions of the antennas used to excite microwaves inside the resonators are indicated by the crosses. The size of the slits is enlarged by a factor of $4$ for their better visibility.}
\label{fig:ExpSetupSketch}
\end{center}
\end{figure}

The principle of the construction of the microwave billiards is illustrated in \reffig{sfig:ExpSetup3d}, which exemplifies a billiard of rectangular shape. They are composed of three plates, $5$ mm thick, made of copper as depicted in \reffig{sfig:ExpSetup3d}. The middle plate has a hole in the form of the shape of the billiard. This yields a microwave cavity with a height of $5$ mm. One of its sides [bottom side in \reffig{sfig:ExpSetup2d}] is composed of three modular bars of copper enabling the variation of the slits size $\slit$ and distance $\dist$ (see also \reffig{fig:SetupGrid}) between the slits, respectively. Microwaves are excited inside the resonator via an antenna located at different positions (crosses). The interference pattern resulting from microwaves leaving the billiard via the slits is recorded on a ``screen,'' which consists of a second antenna moving in the exterior (see below). In order to investigate the relation between the dynamics of the corresponding classical billiard and the interference patterns, a regular billiard with rectangular form and a chaotic one with the shape of a desymmetrized tilted stadium, in the following simply called stadium, were used in the experiments. The three plates are screwed together and wires of solder are inserted into grooves along the contour of the billiard [not shown in \reffig{sfig:ExpSetup3d}] in order to improve the electrical contact between the plates. The rectangular microwave billiard has dimensions \mbox{$768 \times 475$ mm}. The ratio of the side lengths is taken close to the golden ratio $(\sqrt5+1)/2$ in order to avoid degeneracies of the eigenmodes. The stadium billiard is composed of a quarter circle and a trapezoid \cite{Primack1994}. The radius of the quarter circle is $431$ mm, the length of the left side of the stadium is $347$ mm, and the length of the bottom side $872$ mm.

\subsection{\label{ssec:ExpMeas}Measurements}

\begin{figure}[tb]
\begin{center}
\includegraphics[width = 8 cm]{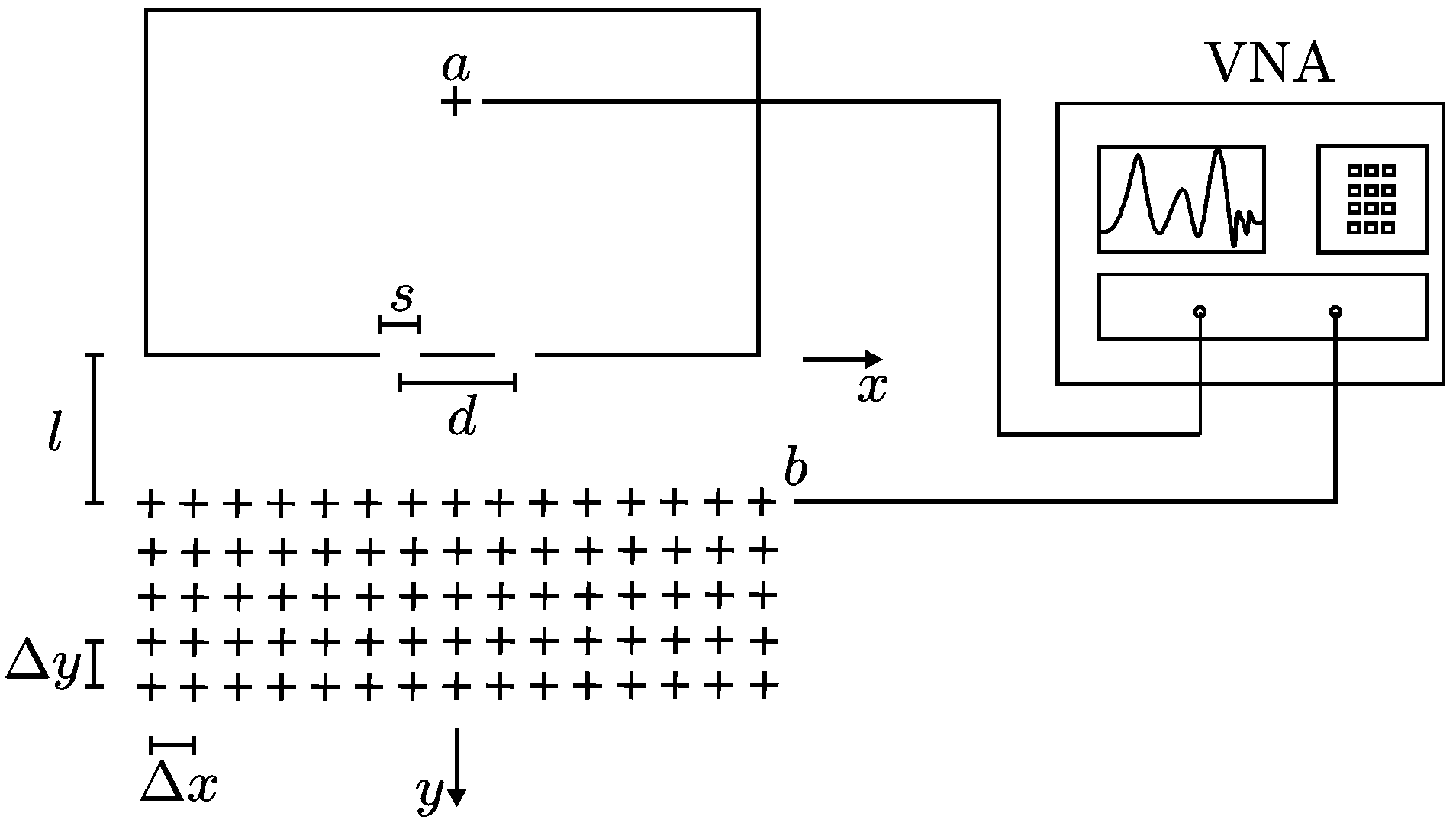} 
\caption{Schematic view (not to scale) of the experimental setup for the measurement of interference patterns. The antenna $a$ is fixed inside the billiard and the antenna $b$ is moved in small steps ($\Delta x$ = $\Delta y$ = 5 mm) in the vicinity of the slits on a grid (crosses) to measure the electric field strengths. The distance between the midpoints of the slits $\dist$ and the slit size $\slit$ are indicated. The vertical distance between the billiard edge and the lines of measuring points parallel to the lower billiard edge is denoted by $l$. The coordinate system \mbox{($x$, $y$)} used in the following is indicated. Its point of origin is the midpoint between the slits. The microwave power is generated and measured, respectively, by the VNA.}
\label{fig:SetupGrid}
\end{center}
\end{figure}

The microwave power is coupled into and out of the setup with wire antennas. They consist of a straight thin wire of about \mbox{0.5 mm} diameter that is soldered in a holder. The electromagnetic signal is generated by a vectorial network analyzer (VNA) of type PNA-L N5230A by Agilent Technologies. It is led to one antenna (port $a$) via a coaxial cable and fed into the cavity. This antenna penetrates partially into the cavity via a hole in the top plate. The output signal is received by another antenna (port $b$) in front of the slits outside the billiards. The ratio of the output power ($P_{\mathrm{out}, b}$) and the input power ($P_{\mathrm{in},a}$) as well as the relative phase of both signals are measured. They yield the complex-valued scattering ($S$)-matrix elements. The squared modulus of the scattering matrix is given as
\begin{equation} \label{ScaMatI} |S_{ba}(f)|^2 = \frac{P_{\mathrm{out},b}(f)}{P_{\mathrm{in},a}(f)} \, . \end{equation}
A graphical representation of $|S_{ba}(f)|^2$ versus the excitation frequency is referred to as a frequency spectrum. Resonance frequencies of the cavity are obtained from the locations of its peaks i.e.\ resonances. Close to the resonance frequency $f_n$ of the $n$th isolated resonance, the $S$-matrix element $S_{ba}$ can be expressed in good approximation \cite{Alt1996, Beck2003} as
\begin{equation} \label{GreFun} S_{ba} \propto \delta_{ba} - i \frac{\gamma_{n, b} \gamma_{n, a}}{f - f_n + i \frac{\Gamma_n}{2}} \, , \end{equation}
where $\Gamma_n$ is the full width at half maximum of the resonance. The numerator $\gamma_{n, b} \gamma_{n, a}$ is proportional to the product of the electric field strengths at the positions $\vec{r}_{a, b}$ of the antennas, i.e.\
\begin{equation} S_{ba}(f_n) \propto E_z(\vec{r}_b) \, E_z(\vec{r}_a) \, . \end{equation}
Accordingly, the electric field distribution $E_z(\vec{r}_b)$ may be determined \cite{StoeckmannBuch2000} by fixing the position of antenna $a$, varying the position $\vec{r}_b$ of antenna $b$ and measuring $S_{ba}$ for each position of antenna $b$, see below in \refsec{stationary}. The electric field intensity $I$ at antenna $b$ is proportional to $|E_z(\vec{r}_b)|^2$ and thus to $|S_{ba}|^2$. 

Figure \ref{fig:SetupGrid} illustrates schematically the experimental setup. Microwave power is coupled into the billiard through \mbox{antenna $a$}, which penetrates $4$ mm into the cavity. In order to measure the leakage of microwave power through the slits to the exterior, the receiving antenna $b$ of $15$ mm length is moved outside of the billiard close to its edge with the slits (see grid of positions in \reffig{fig:SetupGrid}) in steps of \mbox{$5$ mm} which is one third of the minimal considered wavelength. It is connected to the VNA via a flexible cable. The distance $l$ between the billiard edge and the lines of measuring points parallel to the edge (in the following referred to as the screen) varies between $20$ and $500$ mm.

\begin{figure}[tb]
\begin{center}
\includegraphics[width = 8 cm]{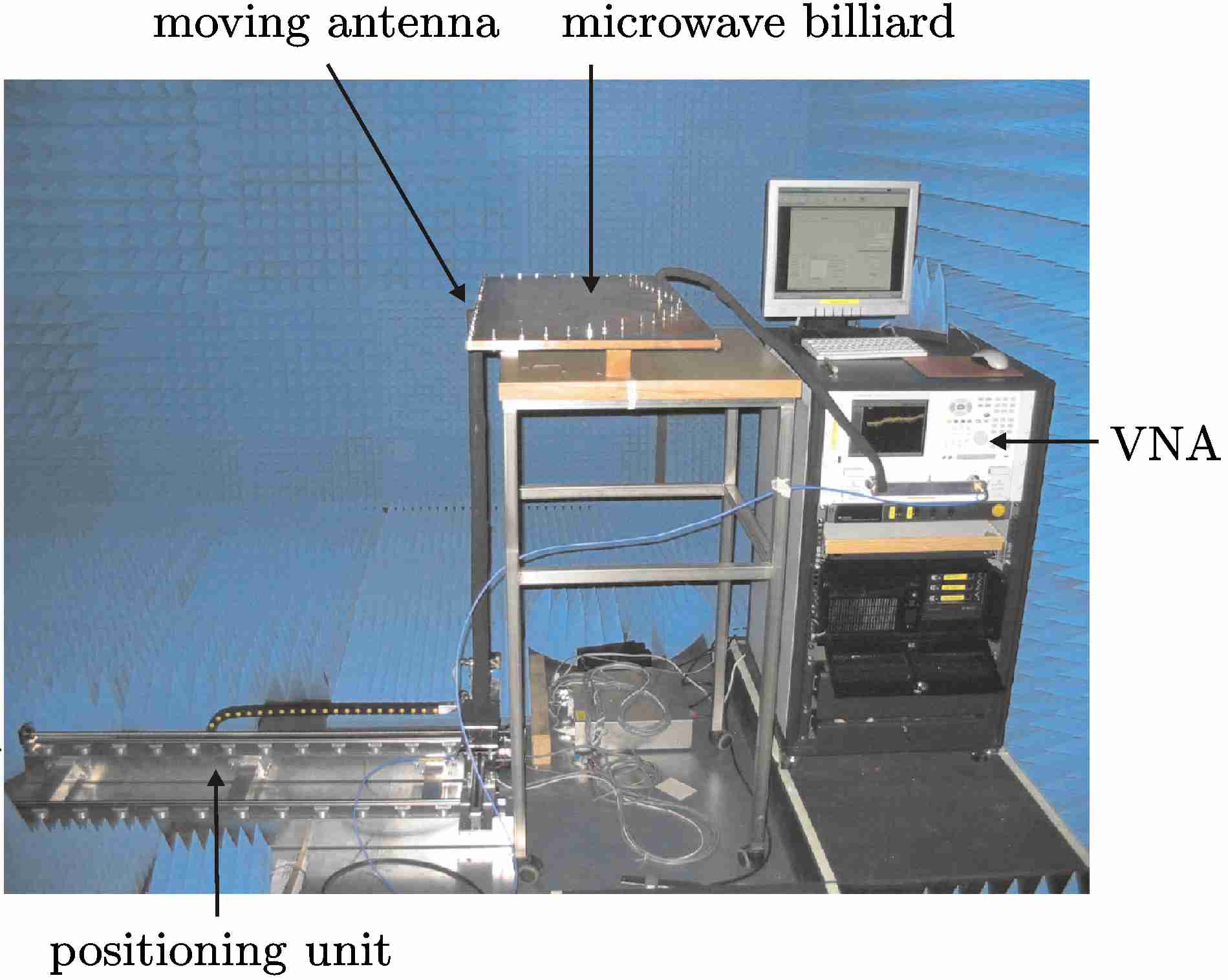} 
\caption{(Color online) Photograph of the measurement setup in the anechoic chamber. The microwave billiard is placed on a table and the receiving antenna is carefully aligned perpendicularly to it. The cable connected to the receiving antenna is coated by a cylindrical double layer of absorber and guided by a positioning unit. The microwave power is coupled from the vectorial network analyzer through a coaxial cable into the cavity.}
\label{fig:AnechChamb}
\end{center}
\end{figure}

\begin{figure}[tb]
\begin{center}
\includegraphics[width = 8 cm]{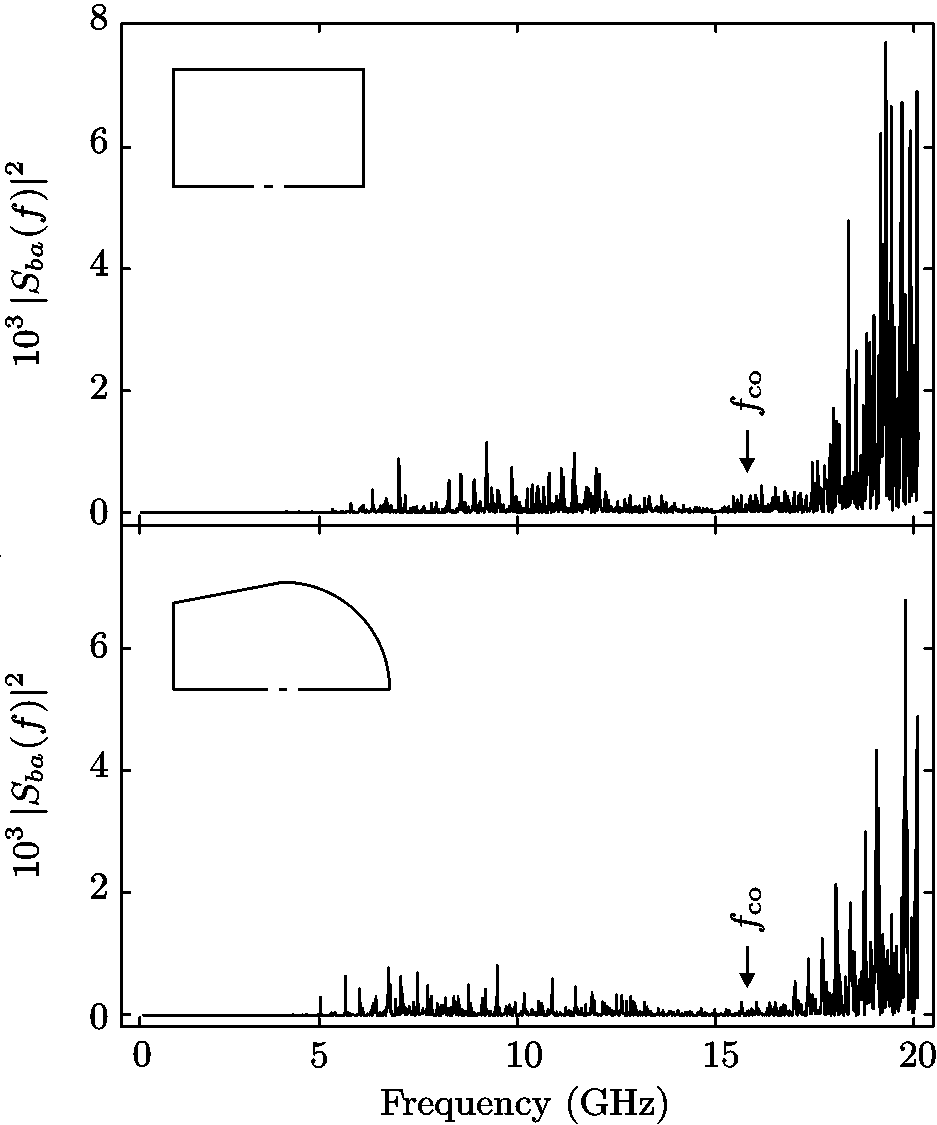} 
\caption{Frequency spectrum of the open two-slit rectangular (top) and stadium (bottom) billiard measured with the emitting antenna $a$ inside the resonator at the point \mbox{$x = 0$ mm}, \mbox{$y = -400$ mm} and the receiving antenna $b$ placed outside at the point $x=0$, \mbox{$y=l=320$ mm}. The slit size is \mbox{$\slit = 9.5$ mm} and the distance between the slits equals $\dist = 78$ mm. The cut-off frequency $f_{co}=c / (2 \slit) = 15.79$ GHz is determined by the slit size and marked by an arrow in both panels.}
\label{fig:FreqSpec}
\end{center}
\end{figure}

The measurements were carried out in a so-called anechoic chamber (see \reffig{fig:AnechChamb}). Pyramidal polyurethane foam structures \mbox{VHP-12 NRL} (from Emerson \& Cuming) cover the whole surface of the chamber and ensure attenuation of the reflected microwave power by $-50$ dB over the whole frequency range of the measurements \cite{EmersonCuming}. Antenna $b$ is guided by a carriage attached to a positioning unit. This unit is driven by bipolar step motors and connected to a computerized numerical control (CNC) module. A PC program communicates with the VNA and with the CNC, enabling the control of the positioning unit and the simultaneous gathering of data. Antenna $b$ is aligned perpendicularly to the billiard plane to ensure that only the $z$ component of the electric field is detected. To prevent a change of the polarization of the electromagnetic field when it escapes into the three-dimensional free space, and to avoid parasitic reflections between the metallic parts of the measurement device and the billiard, all components are covered with microwave absorption material that consists of urethane foam sheets impregnated with carbon (EPP-51 material from ARC Technologies \cite{ARCtech}). Moreover, the copper cable connected to the receiving antenna is coated with an absorptive double layer consisting of urethane foam \mbox{EPP-51} and urethane flat plate EPF-11. The VNA is calibrated in order to remove the effects of the cables and the connectors on the measured spectra. These effects include reflections at connectors, attenuation in the cables and the time delay of the signal accumulated during its passage through the cables. For each position of antenna $b$, the complex $S$-matrix element $S_{ba}$ is measured between antennas $a$ and $b$ from $0.5$ to $20$ GHz with a frequency step of $5$ or \mbox{$10$ MHz}. As a consequence the measurement of the frequency spectrum takes a few seconds for each of the $20\ 000$ positions of the receiving antenna. 

\mbox{Figure \ref{fig:FreqSpec}} shows a frequency spectrum for the rectangular (top) and the stadium (bottom) billiard. The emitting antenna $a$ was positioned inside the resonator symmetrically with respect to the slits at a distance \mbox{$400$ mm} from the lower billiard edge and the receiving antenna $b$ outside at the point $x = 0$, \mbox{$y = l = 320$ mm} (see coordinate system introduced in \reffig{fig:SetupGrid}). In both billiards the slit size $\slit$ is set to $9.5$ mm and the distance $\dist$ between the slits is $78$ mm. The spectrum shows many sharp peaks at resonance frequencies corresponding to quasibound modes. Below the cut-off frequency \mbox{$f_{\rm co}=c / (2 \slit) = 15.79$} GHz defined by the width $\slit$ of the slits, which have the effect of a short waveguide (see \reffig{fig:ExpSetupSketch}) on microwaves leaving the billiard, the transmission amplitude is notably suppressed. Here, $c$ denotes the velocity of light. Note that, due to the short length of the waveguides, the cut-off frequency provides only an approximate value for the onset of the leakage of microwave power through the slits to the exterior.

\section{\label{stationary}Stationary patterns of the intensity on the screen}

\begin{figure}[bt]
\begin{center}
\includegraphics[width = 8 cm]{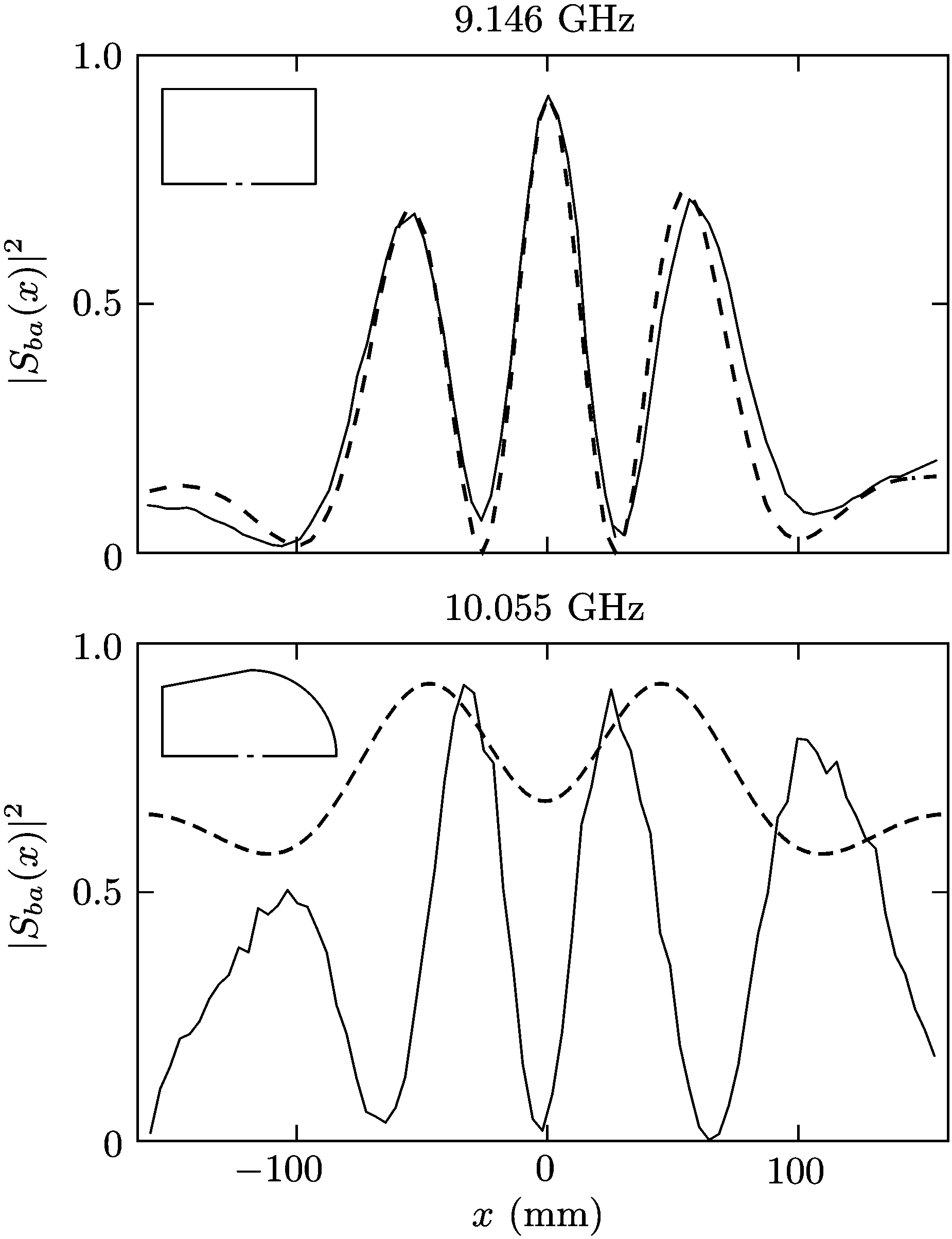} 
\caption{Intensity on the screen (solid line) for the resonance frequency indicated at the top of the panel for the rectangular billiard (top panel) and the stadium billiard (bottom panel). The position of the emitting antenna is chosen on the line $x=0$ at $y = -400$ mm. The screen is located at $l = 155$ mm. The slit sizes and their distance are $\slit = 9.5$ mm and $\dist = 78$ mm, respectively. The dashed line in the top panel results from the Fraunhofer formula [\refeq{FraEqu}] and the one in the bottom panel from the formula in \refeq{BesEqu}.}
\label{fig:StatPattScreen}
\end{center}
\end{figure}

In \reffig{fig:StatPattScreen} we present the squared modulus of the measured transmission scattering matrix element $|S_{ba}|^2$ at a resonance frequency indicated at the top of the respective panels as function of the position $x$ of antenna $b$ on the screen (solid lines) for the rectangular (top panel) and the stadium billiard (bottom panel). The screen is located at a distance \mbox{$l = 155$ mm} from the bottom edge of the billiard. The slit size and the distance between the slits are $\slit = 9.5$ mm and $\dist = 78$ mm, respectively. The position of the emitting antenna is symmetrical with respect to the slits, i.e.\ $x = 0$. For the rectangular billiard, we observe an interference pattern with a central maximum located at $x=0$ and two symmetrical lateral maxima. If the emitting antenna is placed nonsymmetrically with respect to the slits, we obtain a very similar interference pattern, i.e., the measured patterns are nearly independent of the position of the emitting antenna as expected for a stationary field distribution inside the resonator. It should be noted that such interference patterns are observed only at frequencies of sharp resonances. The interference pattern for the rectangular billiard is well described by the Fraunhofer formula,
\begin{equation} \label{FraEqu} I(x) = I_1(x) + I_2(x) + 2 \sqrt{I_1(x) I_2(x)} \cos{(k \dist x / l)} \, , \end{equation} 
which is valid in the far field region if $\slit^2/(l \lambda) \ll 1$ is fulfilled \cite{Hecht2002}. For the parameters of the setup presented in \reffig{fig:StatPattScreen}, $\slit^2/(l \lambda)\approx 0.01$. Here,  $I_1(x)$ and $I_2(x)$ are the intensities radiated by each individual slit and the oscillatory term describes the interferences between the waves emitted through the slits \cite{Hecht2002}. Experimentally, $I_1 (x)$ and $I_2 (x)$ are obtained from measurements with, respectively, one of the slits closed. They are well described by $I(\tilde{x})=\sin^2 (\tilde{x}) / \tilde{x}^2$, where $\tilde{x} = k \slit x / l$. The curve obtained by inserting these intensities into \refeq{FraEqu} is plotted as dashed line in the upper panel of \reffig{fig:StatPattScreen}, showing very good agreement with the measured interference pattern. Similar results of a double-slit experiment with water surface waves instead of microwaves have been published in Ref.\ \cite{Tang2008} where the motion of water waves in a tank with a shallow bottom was investigated. As in the case of microwaves the spatial propagation of the water waves is governed by the two-dimensional Helmholtz equation and is equivalent to the Schr\"odinger equation describing the dynamics in a quantum billiard of corresponding shape.

\begin{figure*}[tb]
\begin{center}
\includegraphics[width = 12 cm]{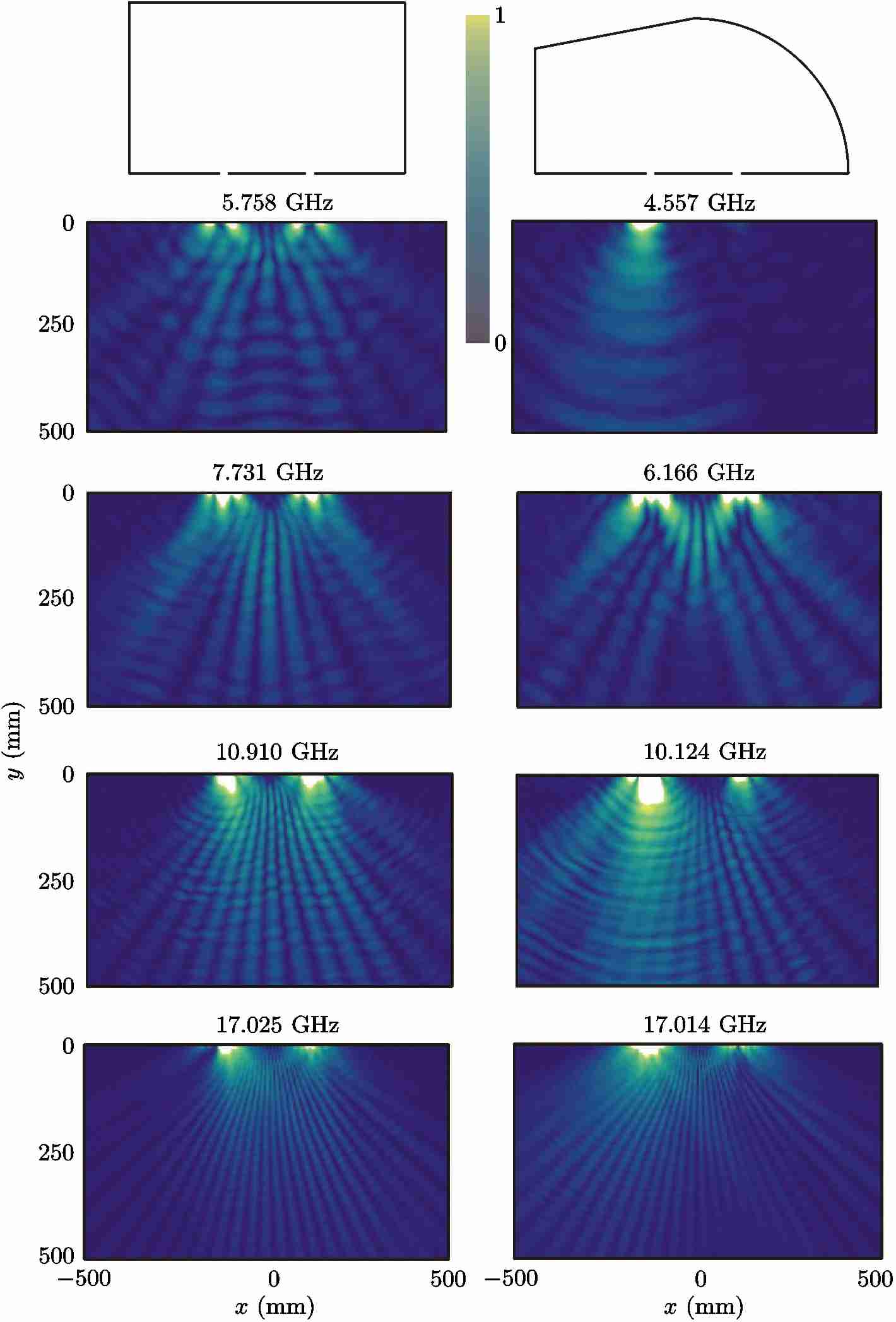} 
\caption{(Color online) Two-dimensional intensity patterns ($I \propto |S_{ba}(x, y)|^2$) on a grid of measuring points for the rectangular (left panels) and stadium billiard (right panels) at a resonance frequency indicated at the top of each panel. Blue (dark) color corresponds to low intensity and yellow (bright) color to high intensity. The slit sizes and their distance are $\slit = 20$ mm and $\dist = 240$ mm, respectively.} 
\label{fig:StatPatt2d}
\end{center}
\end{figure*}

An interference pattern measured with the stadium billiard is shown as solid line in the bottom panel of \reffig{fig:StatPattScreen}. At $x = 0$ we observe a maximum in the interference pattern obtained for the rectangular billiard, whereas that for the stadium billiard exhibits a minimum. According to Berry \cite{Berry1977}, the wave functions of a generic chaotic billiard can be modeled by a superposition of plane waves with the same wave number $|\vec{k}_n| = k$ but random directions and random amplitudes $a_n$, i.e.,
\begin{equation}
\psi_k(\vec{r})= \sum_{n}a_{n} e^{i \vec{k}_n \cdot \vec{r}},
\end{equation}
yielding a Bessel function of the first kind of order $0$ for the spatial correlator \cite{StoeckmannBuch2000},
\begin{equation} \langle \psi_k(\vec{r}) \psi_k(\vec{r}\,' ) \rangle = J_0 (k |\vec{r}-\vec{r}\,'|) \, . \end{equation}
This implies that in the vicinity of the slits the waves are correlated as $J_0(k \dist)$ such that the interference term in \refeq{FraEqu} is modified \cite{Tang2008}. Accordingly,
\begin{equation} \label{BesEqu} \begin{array}{rcl} I(x) & = & I_1(x) + I_2(x) + \\ & & + 2 \sqrt{I_1(x)I_2(x)} J_0(k \dist) \cos{( k \dist x / l )} \, . \end{array} \end{equation} 
For wave numbers $k$ with $J_0(k \dist) = 0$ the interference term vanishes and the intensity should be equal to the sum of the intensities of the single-slit experiments. The intensity corresponding to \refeq{BesEqu} is plotted as dashed line in the lower panel of \reffig{fig:StatPattScreen}. No agreement between \refeq{BesEqu} and the experimental result is found. For the example shown in \reffig{fig:StatPattScreen}, the visibility $(I_{\mathrm{max}}-I_{\mathrm{min}})/I_{\mathrm{max}}$ of the measured pattern is comparable to that for the rectangular billiard. In general the patterns exhibit only for the rectangular billiard such a clear spatial symmetry. However, for all considered cases the interference structure does not disappear, even not for wave numbers for which the interference term should vanish according to \refeq{BesEqu}. It should be noted also that the results for chaotic billiards with water surface waves \cite{Tang2008} do not provide convincing evidence that the total intensity equals the sum of the intensities for the single-slit experiments as predicted in Ref.\ \cite{Casati2005}. 

Examples of intensity distributions in the plane close to the slits are shown in \reffig{fig:StatPatt2d} for the rectangular (left panels) and stadium (right panels) billiard. The slit size is $\slit = 20$ mm and the distance between the slits is set to $\dist = 240$ mm. Antenna $a$ is positioned at $x = 0$. The grid of the positions of antenna $b$ covers an area of \mbox{$0.5$ m$^2$}. In the upper two left panels the intensity patterns at frequencies \mbox{$f = 5.758$ GHz} and \mbox{$7.731$ GHz} corresponding to isolated resonances are shown. They are symmetric with respect to the line $x=0$, as expected due to the symmetry of the setup. In the lower left panels the intensities for frequencies chosen in the regime of overlapping resonances are shown. We observe a slightly asymmetric interference pattern. This is explicable because several modes, which are symmetric or antisymmetric with respect to the symmetry line, are excited simultaneously with different strengths such that the resulting field distribution has no spatial symmetry. Furthermore, experimental imperfections might also contribute to the observed asymmetry. For the stadium billiard (right panels of \reffig{fig:StatPatt2d}) the observation of non-symmetric interference patterns both for frequencies in the regimes of isolated and of overlapping resonances is attributed to the random field distribution of the corresponding modes. There are even cases where microwave power leaks out through just one slit, as, for instance, for \mbox{$f = 4.557$ GHz} in  \reffig{fig:StatPatt2d}. These results provide the first insight into the effect of the shape of the billiards on the interference patterns. 

In order to analyze in more detail the interference patterns of the waves leaving the billiards we define the quantities
\begin{equation} \label{Corr1} \Delta_\rho = \ln \left\{ \frac{|E_z(x_0+\delta x)|^2}{|E_z(x_0-\delta x)|^2} \right\} \, , \end{equation}
\begin{equation} \label{Corr2} \Delta_\phi = \arg [E_z(x_0+\delta x)]- \arg [E_z(x_0-\delta x)] \, , \end{equation}
which correlate the intensities and the phases, respectively, of the electric field $E_z$ at points $x_0 + \delta x$ and $x_0 - \delta x$, where $x_0$ denotes the central position on the screen. These correlators have been introduced and discussed in Ref.\ \cite{Levnajic2010} and provide information about the symmetry and visibility of the interference structure. In a double-slit experiment with plane waves the phase difference of the exiting waves at the slits determines the symmetry of the obtained pattern whereas the amplitude difference should account for the visibility. If the amplitudes and the phases of the fields $E_z$ at the positions $x_0 + \delta x$ and $x_0 - \delta x$ coincide, i.e. if \mbox{$\Delta_\rho = \Delta_\phi = 0$}, the pattern is symmetric with visibility of $100 \%$.

\begin{figure}[tb]
\begin{center}
\includegraphics[width = 8 cm]{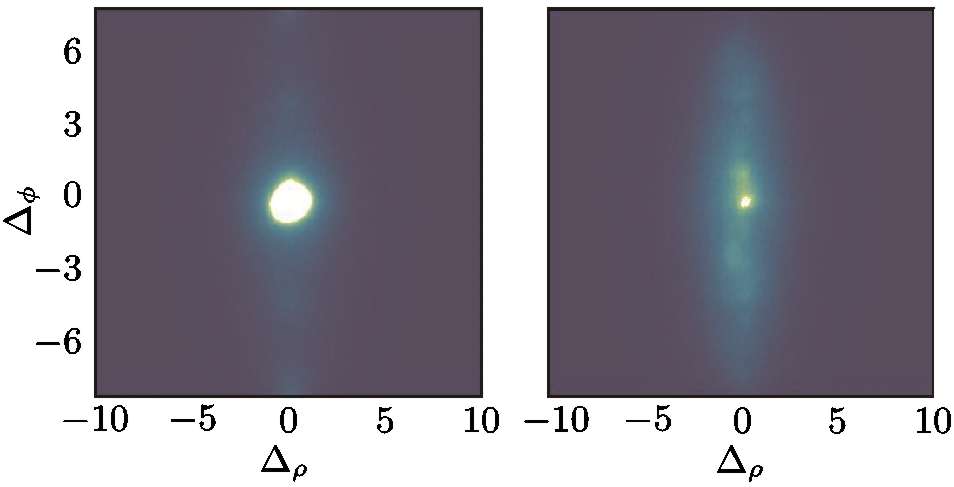} 
\end{center}
\caption{(Color online) Joint probability distributions of the correlators $\Delta_\rho$ and $\Delta_\phi$ of the electric fields defined by \mbox{Eqs.\ (\ref{Corr1})} and (\ref{Corr2}) outside the rectangular billiard (left panel) and the stadium billiard (right panel). Yellow (bright) color corresponds to high and blue (dark) color to low probability. The pairs ($\Delta_\rho$, $\Delta_\phi$) are evaluated for $30$ excitation frequencies on a grid of $10\ 000$ measuring points. The slit sizes are \mbox{$\slit = 20$ mm} and their distance is \mbox{$\dist = 240$ mm} as in \reffig{fig:StatPatt2d}.}
\label{fig:CorrJoin}
\end{figure}

\begin{figure}[tb]
\begin{center}
\includegraphics[width = 8 cm]{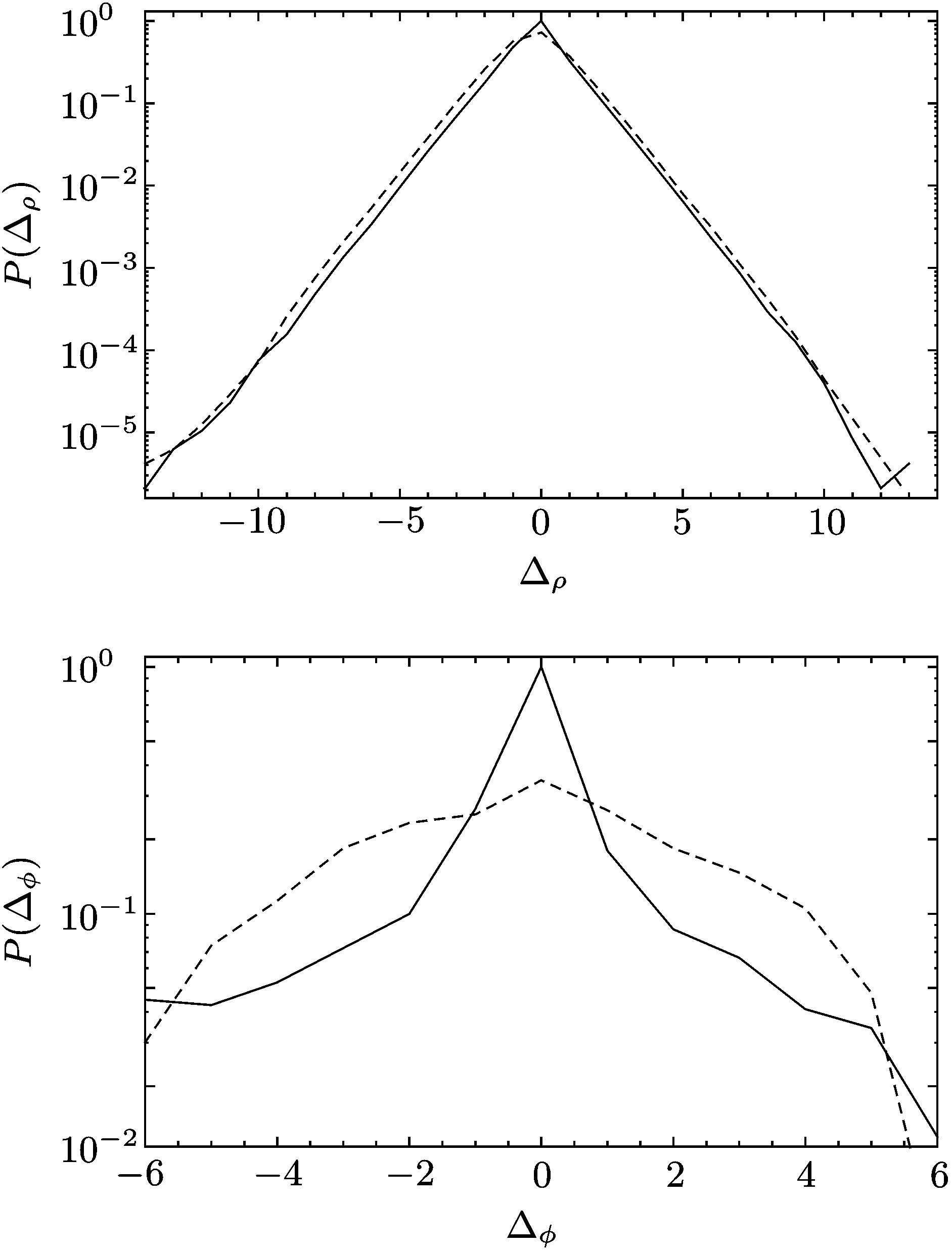} 
\caption{Distributions of the correlators $\Delta_\rho$ (top panel) and $\Delta_\phi$ (bottom panel) for the rectangular (solid lines) and the stadium billiard (dashed lines) in semilogarithmic scale. The data set is the same as that in \reffig{fig:CorrJoin}.} 
\label{fig:CorrDistr}
\end{center}
\end{figure}

These correlators were evaluated for the measurements shown in \reffig{fig:StatPatt2d}. Therefore, $x_0 = 0$ and the screen is located at the bottom side of the billiards at a distance from $5$ mm up to $500$ mm with a spatial resolution of $5$ mm in $y$ and $\delta x$. For each value of $y$ the pair $(\Delta_\rho, \Delta_\phi)$ was determined for $100$ values of $\delta x$ (see \reffig{fig:SetupGrid}), yielding a total number of $10\ 000$ measuring points. The parameters $\slit$ and $\dist$ are $20$ and \mbox{$240$ mm}, respectively. Thirty resonance frequencies of the corresponding closed billiards are chosen as excitation frequencies. Thus we obtain $300\ 000$ pairs of ($\Delta_\rho$, $\Delta_\phi$). The resulting joint  probability distribution of ($\Delta_\rho$, $\Delta_\phi$) is shown in \reffig{fig:CorrJoin}. For the rectangular billiard (left panel), the probability distribution is strongly concentrated around $(\Delta_\rho, \Delta_\phi) = (0, 0)$. This yields an evidence for the high (though not perfect) spatial symmetry of the measured interference patterns (cf.\ the left panels of \reffig{fig:StatPatt2d}). For an ideal double slit experiment with plane waves, one would expect a nonvanishing distribution only at $(0, 0)$. The distribution for the stadium billiard is shown in the right panel. It is also centered around $(0, 0)$, but in comparison to the left panel much less concentrated in the direction of the $\Delta_\phi$ coordinate. This is expected due to the slight asymmetry in the interference patterns of the stadium billiard observed in the right panels of \reffig{fig:StatPatt2d}. In the top and bottom panels of \reffig{fig:CorrDistr} the distributions of, respectively, the intensity and phase correlator are presented. The distributions of $\Delta_\rho$ are almost identical for the rectangular (solid line) and the stadium (dashed line) billiard and decay exponentially with $|\Delta_\rho|$, i.e., the visibility of the measured interference patterns is similar for both billiards (cf.\ \reffig{fig:StatPattScreen}). In contrast, the distributions of $\Delta_\phi$ (bottom panel) differ markedly. While the distribution is peaked at $\Delta_\phi = 0$ for the rectangular billiard, i.e., there most patterns are symmetric, it is bell shaped and much broader around $\Delta_\phi = 0$ for the stadium billiard. As above, this is attributed to the strongly nonsymmetrical patterns caused by the random field distributions as shown in the right panels of \reffig{fig:StatPatt2d}. In conclusion, the distributions of the correlators show a clear difference for the interference patterns of the regular rectangular and the chaotic stadium billiard.

\section{\label{temp}Temporal evolution of the fields outside the billiards}
In this section we analyze and discuss the time evolution of the field in the vicinity of the double slit. The emitting antenna $a$ is placed inside the billiard symmetrically with respect to the slits at a distance of $400$ mm to the edge with the slits whereas the distance $l$ of the screen to this edge equals $320$ mm. Close to an isolated resonance, $S_{ba}$ is essentially given by a modified Green function up to a factor that varies slowly with the frequency \cite{Stein1995}. Consequently, the propagator $K(\vec{r_b},\vec{r_a},t,0)$ is related to the Fourier transform of the experimentally obtained scattering matrix \cite{Schaefer2006}, i.e.,  
\begin{equation} \label{FouTra} \begin{array}{c} K(\vec{r}_b,\vec{r}_a, t, 0) \sim \tilde{S}_{ba}(\vec{r}_b,\vec{r}_a, t, 0) \\ \\ = \int_{-\infty}^{\infty}S_{ba}(f)e^{-2\pi i t f} df \, . \end{array} \end{equation}
The experimental spectra are measured at $N$ discrete frequencies in the range from $f_{\mathrm{min}}$ to $f_{\mathrm{max}}$ and, correspondingly, the discrete Fourier transform is evaluated in \refeq{FouTra}. The maximal time accessible is determined by $1 / \Delta f$, where $\Delta f$ is the frequency step used in the experiment. In the present measurements $\Delta f = 10$ MHz, which corresponds to a maximal time of $100$ ns.

\begin{figure}[bt]
\begin{center}
\includegraphics[width = 8 cm]{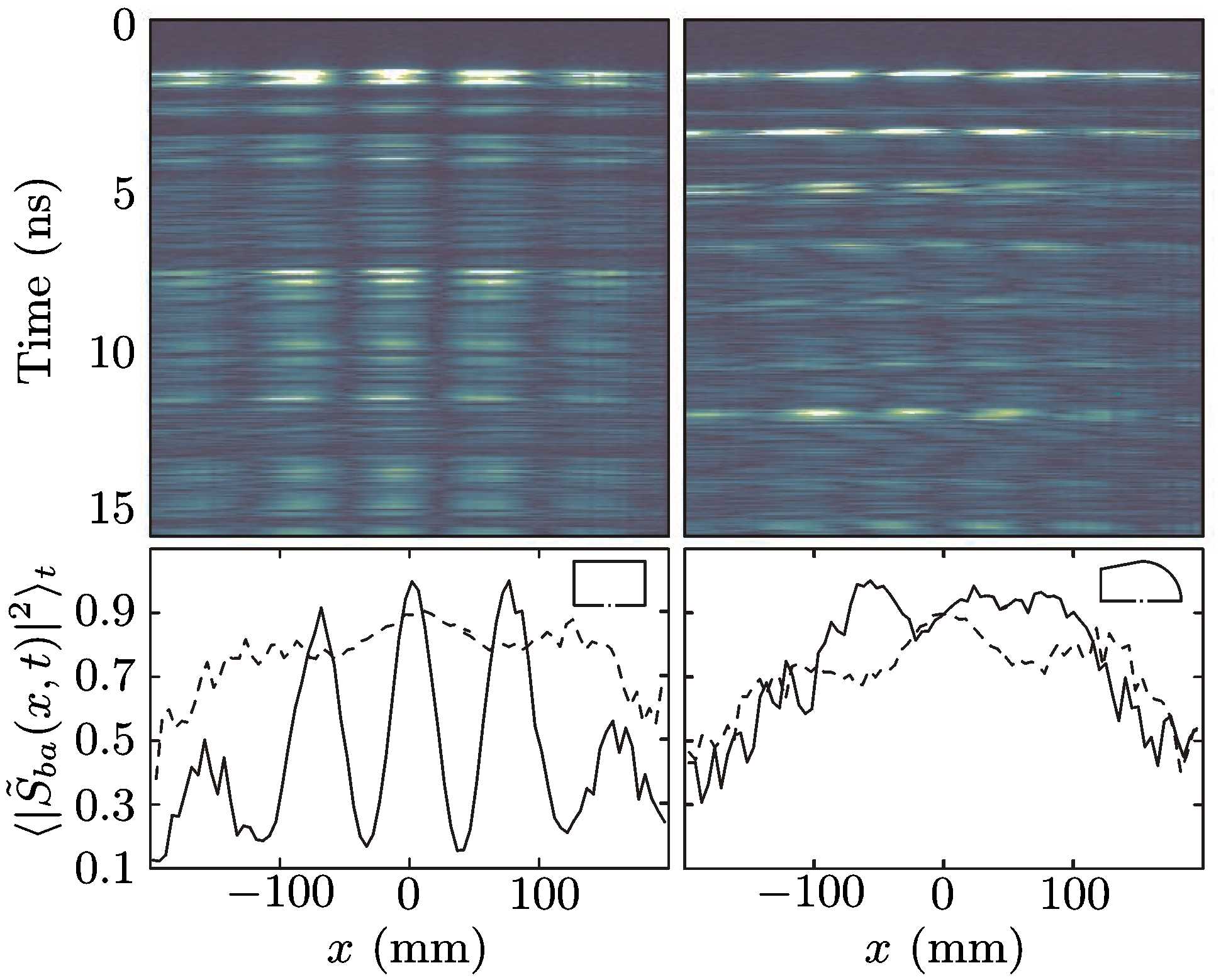} 
\caption{(Color online) Time evolution (upper panels) of the intensity measured on a screen and its average (lower panels) for the rectangular (left panels) and the stadium billiard (right panels). The upper panels and the solid lines in the lower panels result from the double-slit experiment, and the dashed lines in the lower panels show the time-averaged superposition of the intensities (average taken up to $75$ ns) of the two corresponding single-slit experiments. The time evolution is obtained from the measured scattering matrix $S_{ba}(f)$ via a Fourier transform [cf.\ \refeq{FouTra}]. For the color scale, see \reffig{fig:StatPatt2d}. The slit sizes are $\slit = 9.5$ mm and their distance is set to $\dist = 78$ mm. The distance of the screen from the billiard is $l = 320$ mm.}
\label{fig:ScreenTime}
\end{center}
\end{figure}

The upper panels of \reffig{fig:ScreenTime} show the time evolution of the intensity on the screen for the rectangular (left) and the stadium (right) billiard, i.e., $|\tilde{S}_{ba}(x, t)|^2$ versus the position $x$ of antenna $b$. The time a wave emitted from antenna $a$ needs to reach antenna $b$ on the screen is called escape time. At each escape time, an interference pattern is observed as an intensity variation along $x$. It results from the interference between the waves emitted from the two slits. As a result of the spatial symmetry of the rectangular billiard and the choice of the position of the emitting antenna $a$ at $x=0$, every part of the cylindrical wave pulse emitted by this antenna with initial momentum direction $\vec{k} = (k_x, k_y)$ travels the same path as its counterpart with $\vec{k} = (-k_x, k_y)$. This is a consequence of the omnidirectionality of the initial wave pulse. Accordingly, these waves hit the slits with no phase difference. The positions of the maxima and minima observed in the intensity pattern in the upper left panel of \reffig{fig:ScreenTime} do not change with time. For the stadium billiard, however, there is no spatial symmetry and, consequently, the positions of the maxima and minima in the upper right part of  \reffig{fig:ScreenTime} change with time. 

In order to quantify how the interference patterns are influenced by the time evolution of the waves emitted from the two slits, for each position $x$ of antenna $b$ the intensities $|\tilde{S}_{ba}(x, t)|^2$ determined at discrete times $t_i$ are averaged over time up to $t_{N} = 75$~ns, yielding $\langle |\tilde{S}_{ba}(x, t)|^2 \rangle_t=\frac{1}{N}\sum_{i=1}^{N}|\tilde{S}_{ba}(x, t_i)|^2$. It should be noted that this quantity corresponds to the averaged current evaluated in the numerical simulation by Casati and Prosen~\cite{Casati2005}. The time average is performed up to $75$ ns since $|\tilde{S}_{ba}(x, t)|^2$ is negligibly small for longer times. The resulting time-averaged intensities are shown in the lower panels of \reffig{fig:ScreenTime} as solid lines. For the rectangular billiard the interference structure is still clearly visible, whereas it disappears for the stadium billiard. In addition, the sum of the time-averaged intensities $\langle |\tilde{S}_{ba}(x, t)|^2 \rangle_t$ obtained from two single-slit experiments in which the left, respectively, right slit was closed (dashed lines) is shown for comparison. It has no interference structure and shows a qualitative agreement with the time-averaged intensity resulting from the double-slit experiment only for the fully chaotic stadium billiard.

If the emitting antenna is not placed on the symmetry line, the interference pattern of the time-averaged intensity distribution disappears also for the rectangular billiard and resembles that for the stadium cavity. Indeed, waves emitted with initial wave vector $\vec{k} = (k_x, k_y)$ and $\vec{k} = (-k_x, k_y)$ do not cover the same path length until they reach the slits and thus they have a phase relation varying in time. This finding confirms that the mechanism leading to the emergence of the interference patterns observed in the left panels of \reffig{fig:ScreenTime} is not linked to the classical dynamics of the billiard but to spatial symmetry properties of the system consisting of the billiard with the slits and the emitting antenna. We may expect from these results that no interference patterns are observed when the slits are not positioned symmetrically. This was not tested experimentally, though. Only recently our attention was drawn to a work by Fonte and Zerbo, where similar results as observed in our measurements were obtained with numerical simulations \cite{Fonte2006}. 

We also performed ray-tracing simulations, where classical pointlike particles were injected from the position of antenna $a$ with the angles of initial direction sweeping the whole interval from $0$ to $2\pi$. The time of flight $t_f$ the particles need to reach the position of antenna $b$ was computed. We observe a good agreement between the measured time spectra $|\tilde{S}_{ba}(x, t)|^2$ and the classical ones up to a time $t^*\approx 15$~ns which correponds to a length of $ct^*\approx 4.5$ m. The latter are obtained by counting the number of particles that need a certain time of flight and plotting it versus $t_f$. 

The results of the experiments with a single emitting antenna presented in this section thus lead to the following conclusions:
\begin{itemize}

\item Clear interference patterns that are symmetric with respect to the two slits are observed only for the rectangular billiard with the emitting antenna placed on the symmetry line $x = 0$. These are well described by the Fraunhofer formula, \refeq{FraEqu}. Furthermore, the time averaged intensities show clear interference patterns for the rectangular billiard if the emitting antenna is positioned symmetrically.

\item Interference patterns are not observed in the time-averaged intensities resulting from the stadium billiard and also not for the rectangular billiard with the emitting antenna placed asymmetrically with respect to the slits.

\item For the stadium billiard, the sum of the time-averaged intensities of the two single-slit experiments is roughly equal to that of the double-slit experiment.   

\end{itemize}

The experiments presented in this section were performed with an omnidirectional wave pulse emitted from a single antenna. However, 
as mentioned in \refsec{introduction}, the initial states considered in the numerical simulations presented in Ref.\ \cite{Casati2005} are localized wave packets with a well-defined direction. For the realization of a similar experimental situation we developed a method for the construction of directional wave packets using an array of antennas, which is presented in the following section. It was used recently in Ref.\ \cite{Unterhinninghofen2011} to generate plane waves from a superposition of cylindrical waves.

\section{\label{gaussian}Experiments with a directional wave packet}

In this section we present a new method for the construction of directional wave packets. Their time evolution is used for the analysis of the experimental spectra in order to study the dependence of the interference patterns on the billiard geometry as well as on the initial angle. An arbitrary initial state $E_z(\vec{r}_0,t_0)$ evolves according to 
\begin{equation} \label{GauStaS} \begin{array}{rcl} E_z(\vec{r},t) & \propto & \int_{\mathcal{G}} d^2 \vec{r}_0 K (\vec{r},\vec{r}_0,t,t_0)  \frac{\partial E_z(\vec{r}_0,t_0)}{\partial t} \, - \\ \\ & & \int_{\mathcal{G}} d^2 \vec{r}_0 \frac{\partial K(\vec{r},\vec{r}_0,t,t_0)}{\partial t}  E_z(\vec{r}_0,t_0) \, , \end{array} \end{equation} 
where $K(\vec{r},\vec{r}_0,t,t_0)$ is the electromagnetic propagator between two antennas at positions $\vec{r}_0$ and $\vec{r}$ \cite{Morse1953}. The second term in \refeq{GauStaS} can be set to zero by a proper choice of initial conditions, i.e., $E_z (\vec{r}_0, t_0) = 0$. For the analysis of the experimental data we replace $K(\vec{r},\vec{r}_0,t,t_0)$ by the Fourier transform  $\tilde{S}_{ba}(\vec{r},\vec{r}_0,t,t_0)$  of the measured transmission scattering matrix element [see \refeq{FouTra}]. Here $\vec{r}$ denotes the position of the receiving antenna $b$ and $\vec{r}_0$ that of the emitting antenna $a$. Experimentally, we can obtain $\tilde{S}_{ba}(\vec{r},\vec{r}_0,t,t_0)$ only for discrete positions $\vec{r}_{0i}$ of the emitting antenna. Accordingly, the integral entering \refeq{GauStaS} has to be replaced by a sum,
\begin{equation} \label{GauSta2} E_z(\vec{r},t) \propto \sum_i \tilde{S}_{ba}(\vec{r},\vec{r}_{0i},t,t_0) \frac{\partial E_z(\vec{r}_{0i},t_0=0)}{\partial t} \, .  \end{equation}
\begin{widetext}
In analogy to Ref.\ \cite{Casati2005}, we set
\begin{equation} \label{GauSta1der} \frac{\partial E_z}{\partial t}(\vec{r}_0,0) = -i \omega E_0 \exp\left( -\frac{(\vec{r}_0-\vec{R})^2}{2\sigma^2} \right) \exp[i \vec{k} \cdot (\vec{r}_0-\vec{R})] \, , \end{equation}
where $\vec{R}$ denotes the central position of the initial wave packet, $\sigma$ its width and $\vec{k}$ the wave vector. Inserting \refeq{GauSta1der} into \refeq{GauSta2} we obtain
\begin{equation} \label{GauSta3} \EzWP(\vec{r},t) \propto - i \omega E_0 \sum_i \exp \left( -\frac{(\vec{r}_{0i}-\vec{R})^2}{2\sigma^2} \right) \exp[i \vec{k} \cdot (\vec{r}_{0i}-\vec{R})] \tilde{S}_{ba}(\vec{r},\vec{r}_{0i},t,t_0) \, . \end{equation}
\end{widetext}

\begin{figure}[!b]
\begin{center}
\includegraphics[width = 3 cm]{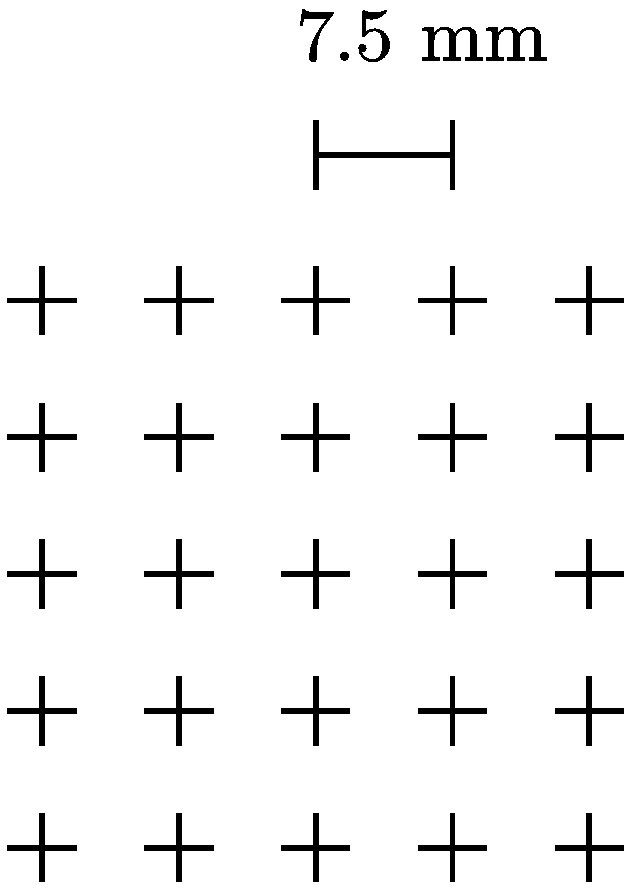}
\caption{Array of $5 \times 5$ positions (crosses) of the emitting antenna for the generation of a directional wave packet in free space. For each of its $25$ positions the field distribution around it is measured with a moving receiving antenna (see text for details).}
\label{fig:freePropSetup}
\end{center}
\end{figure}

Equation (\ref{GauSta3}) is the final expression for the time evolution of the constructed wave packet, $\EzWP(\vec{r},t)$. It is proportional to a sum over the Fourier transforms of the spectra weighted by complex coefficients. The initial direction and the width of the wave packet can be varied arbitrarily since $\vec{k}=k (\cos \alpha,\sin \alpha)$ and $\sigma$ are free parameters. However, the width $\sigma$ should be chosen larger than the minimal wavelength used in the experiments, i.e.\ \mbox{$\sigma > 15$ mm}. On the other hand, if it is chosen to be much larger than the size of the domain of emitting antennas, the coefficients \mbox{$\exp \{ -(\vec{r}_{0i}-\vec{R})^2 / (2\sigma^2)\}$} would all be approximately equal to $1$ such that all antennas would provide the same contribution to the sum in \refeq{GauSta3}. Thus $\sigma$ is chosen of the order of the size of this domain.

\begin{figure}[tb]
\begin{center}
\includegraphics[width = 8 cm]{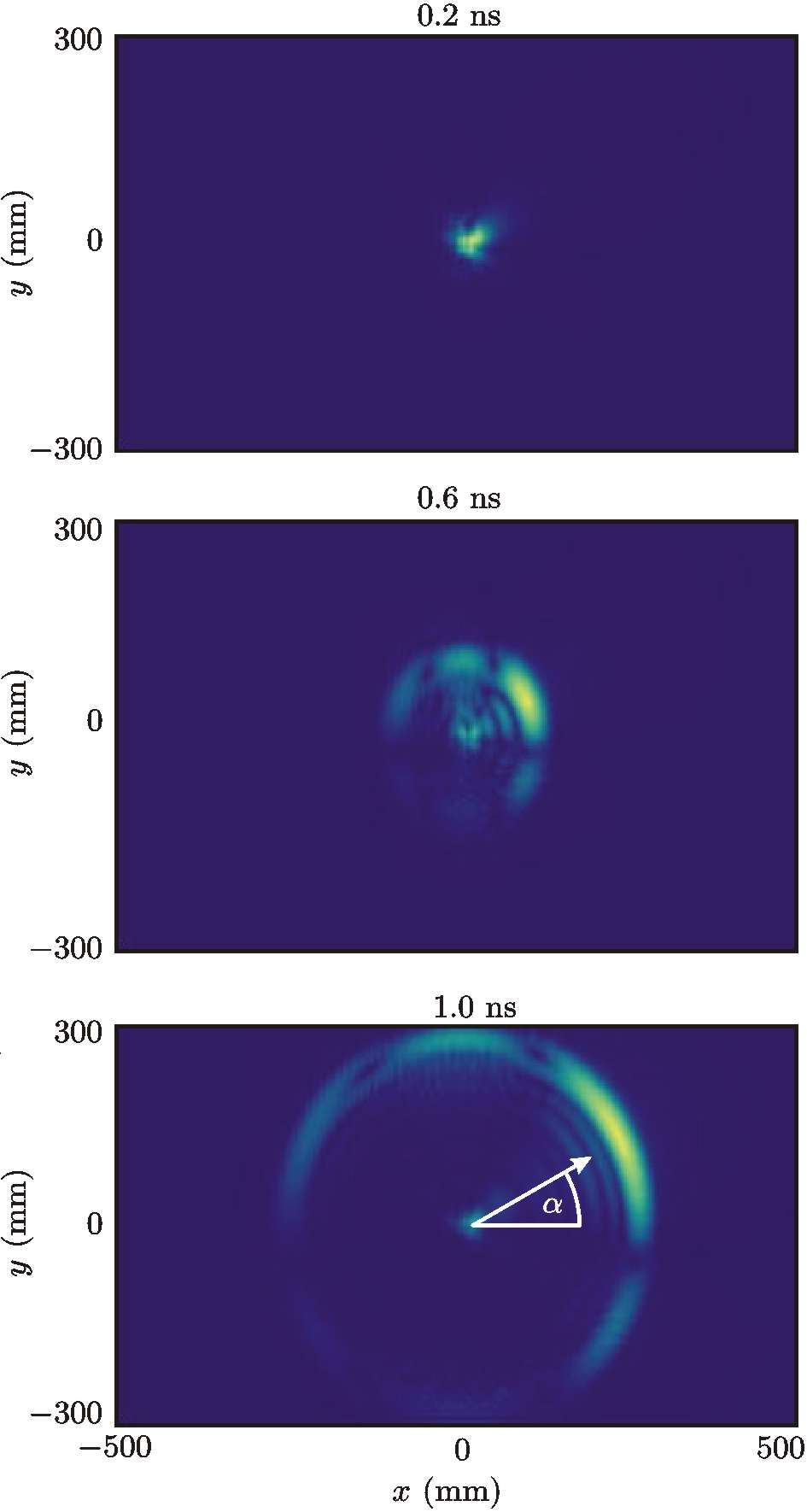} 
\caption{(Color online) Three snapshots of the propagation in free space of a wave packet constructed according to \refeq{GauSta3}, where the time spectra were obtained from measurements for the $25$ positions of the emitting antenna shown in \reffig{fig:freePropSetup} with the center position at $(x, y) = (0, 0)$. The direction of the wave packet is indicated by the angle $\alpha$. For the color scale see \reffig{fig:StatPatt2d}.}
\label{fig:freeProp}
\end{center}
\end{figure}

We tested the propagation of the wave packet constructed according to \refeq{GauSta3} experimentally in free space. A single emitting antenna $a$ hung down from the ceiling of an empty room at a position $\vec{r}_a = \vec{r}_{0i}$ and the position $\vec{r}_b = \vec{r}$ of the receiving antenna was varied in the vicinity of $\vec{r}_{0i}$ in a plane perpendicular to the emitting wire antenna. For each position $\vec{r}$ the transmission spectrum $S_{ba}(\vec{r}, \vec{r}_{0i}, f)$ was measured for frequencies from $0.5$ to $20$ GHz. These transmission measurements were repeated for altogether $25$ positions $\vec{r}_{0i}$ of the emitting antenna $a$ on a square grid (see schematic sketch in \reffig{fig:freePropSetup}). Using the principle of linear superposition, for each position $\vec{r}$ of antenna $b$ the time spectra for the $25$ different positions $\vec{r}_{0i}$ of the emitting antenna are added up according to \refeq{GauSta3}. Three snapshots for the resulting wave packet propagation are shown in \reffig{fig:freeProp}, with the width of the of the wave packet and its initial angle chosen as $\sigma = 20$ mm, respectively, $\alpha = 31^\circ$. As observed in the upper panels the wave packet moves into the expected direction and spreads out fast. However, the spatial structure reveals propagation in additional directions. We attribute this effect to the discretization of the initial positions $\vec{r}_0$ in \refeq{GauStaS}.

\begin{figure}[bt]
\begin{center}
\includegraphics[width = 8 cm]{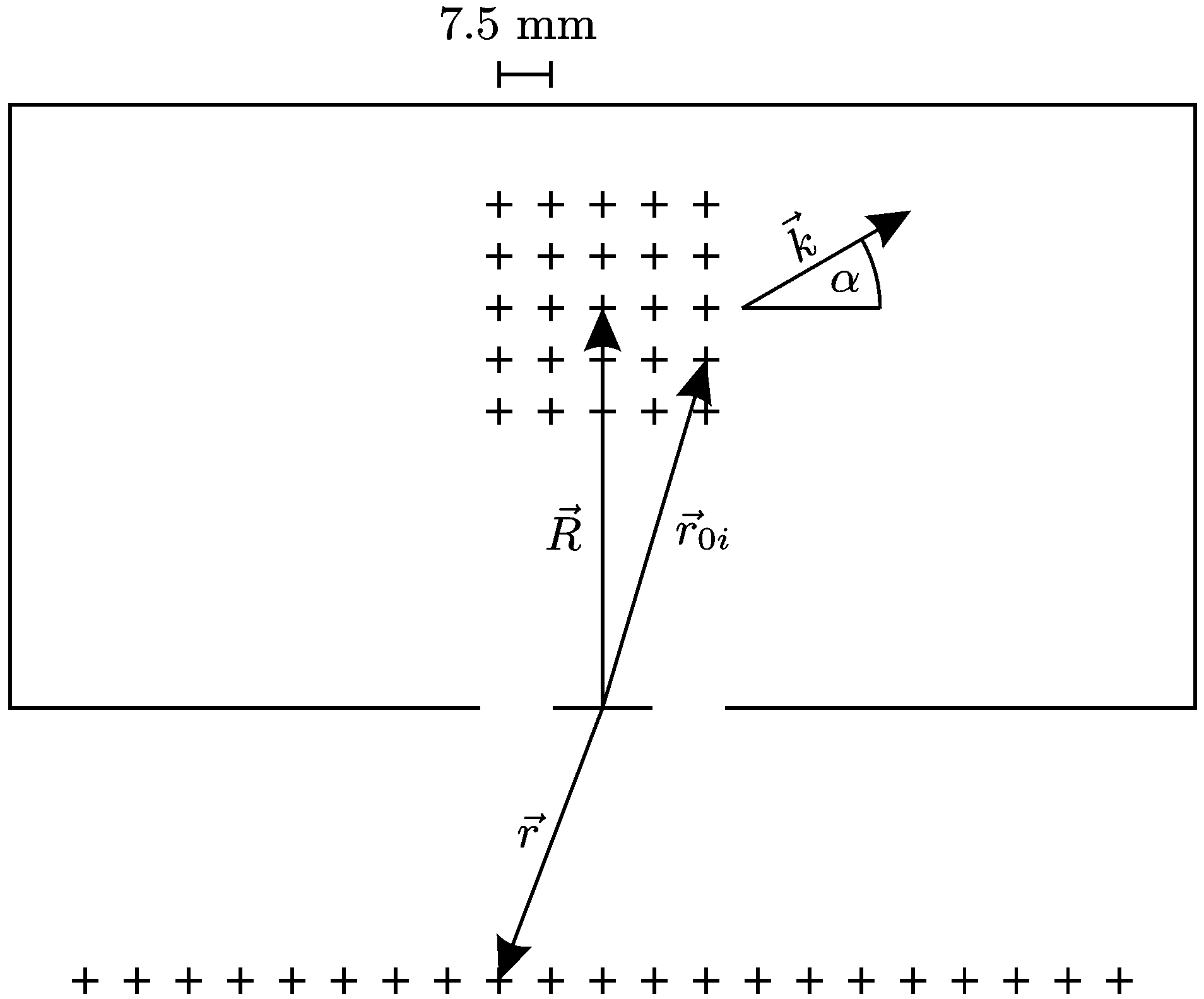} 
\caption{Sketch of the experimental setup for the construction of a directed wave packet inside the resonator (not to scale). An array of $5 \times 5$ holes for the emitting antenna is drilled into the top billiard plate.}
\label{fig:SetupWP}
\end{center}
\end{figure}

For the double slit experiments \mbox{$5 \times 5$} antenna holes arranged in a square array at distances of \mbox{$7.5$ mm} were drilled into the top plates of the billiards as shown in the sketch of \reffig{fig:SetupWP}. The position of the central antenna of the squared array is located on the line $x=0$ at a distance of $400$ mm from the edge with the slits. In analogy to the measurement in free space presented above, one emitting antenna $a$ is put consecutively into the $25$ holes at positions $\vec{r}_{0i}$ and the field distributions outside the billiards are obtained by measuring frequency spectra with the moving antenna $b$. The computed time spectra are then added according to \refeq{GauSta3}. The initial angle of propagation $\alpha$ of the wave packet constructed this way is indicated in \reffig{fig:SetupWP}. It is chosen such that the wave packet is emitted along periodic orbits of the rectangular billiard.

\begin{figure*}[tb]
\begin{center}
\includegraphics[width = 12 cm]{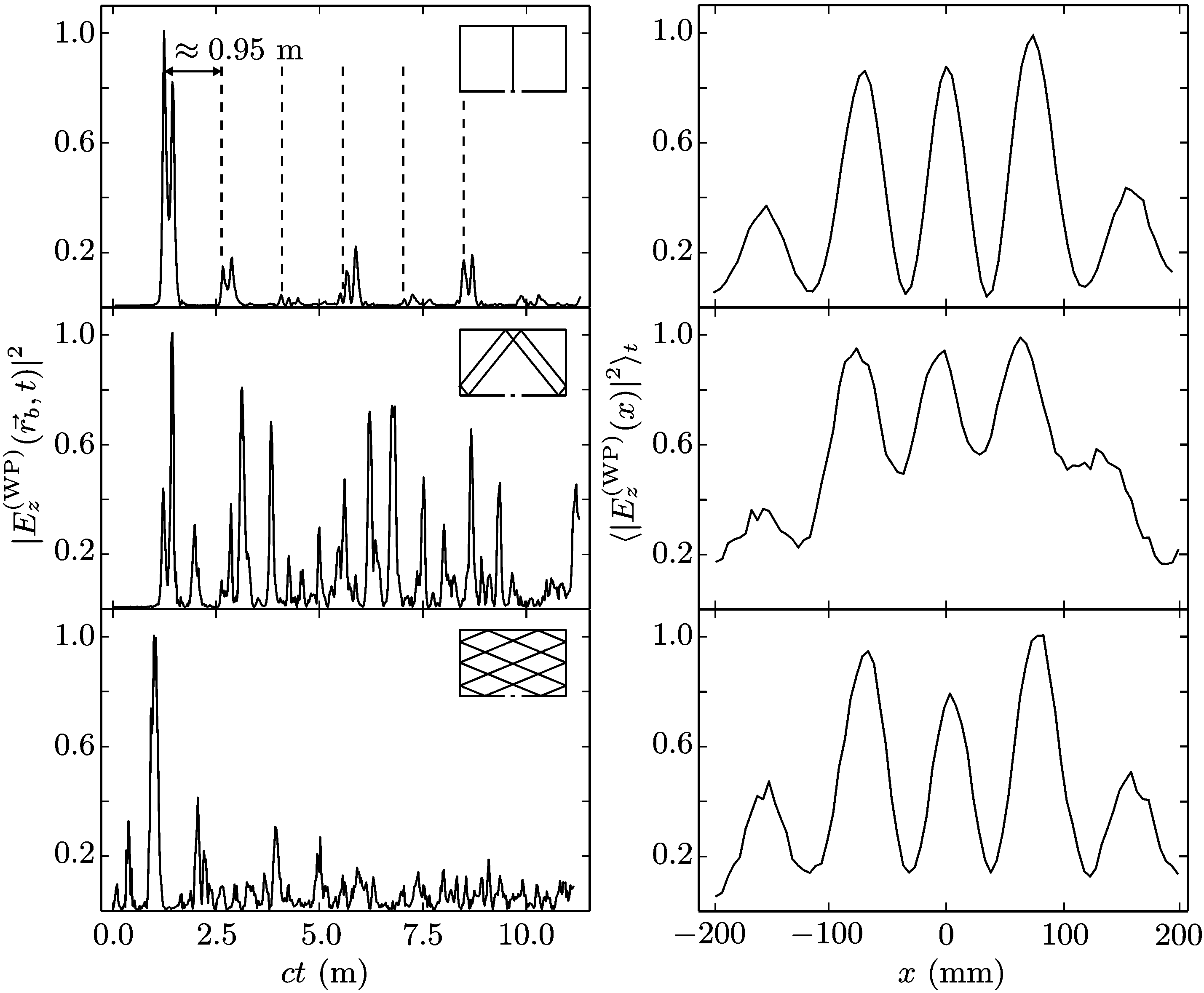} 
\caption{Time spectra (left panels) with antenna $b$ at $x = 0$ and time-averaged intensity (right panels) for a wave packet constructed according to \refeq{GauSta3} which starts with an initial angle of $270^\circ$ (top), $51^\circ$ (middle), and $22^\circ$ (bottom) in the rectangular billiard. The periodic orbits corresponding to these angles are shown as insets. The dashed lines in the top left panel indicate the periodicity of the escape times. The parameters $\dist$, $\slit$, and $l$ are the same as in \reffig{fig:ScreenTime}.}
\label{fig:EchoesWpRect}
\end{center}
\end{figure*}

\begin{figure*}[tb]
\begin{center}
\includegraphics[width = 12 cm]{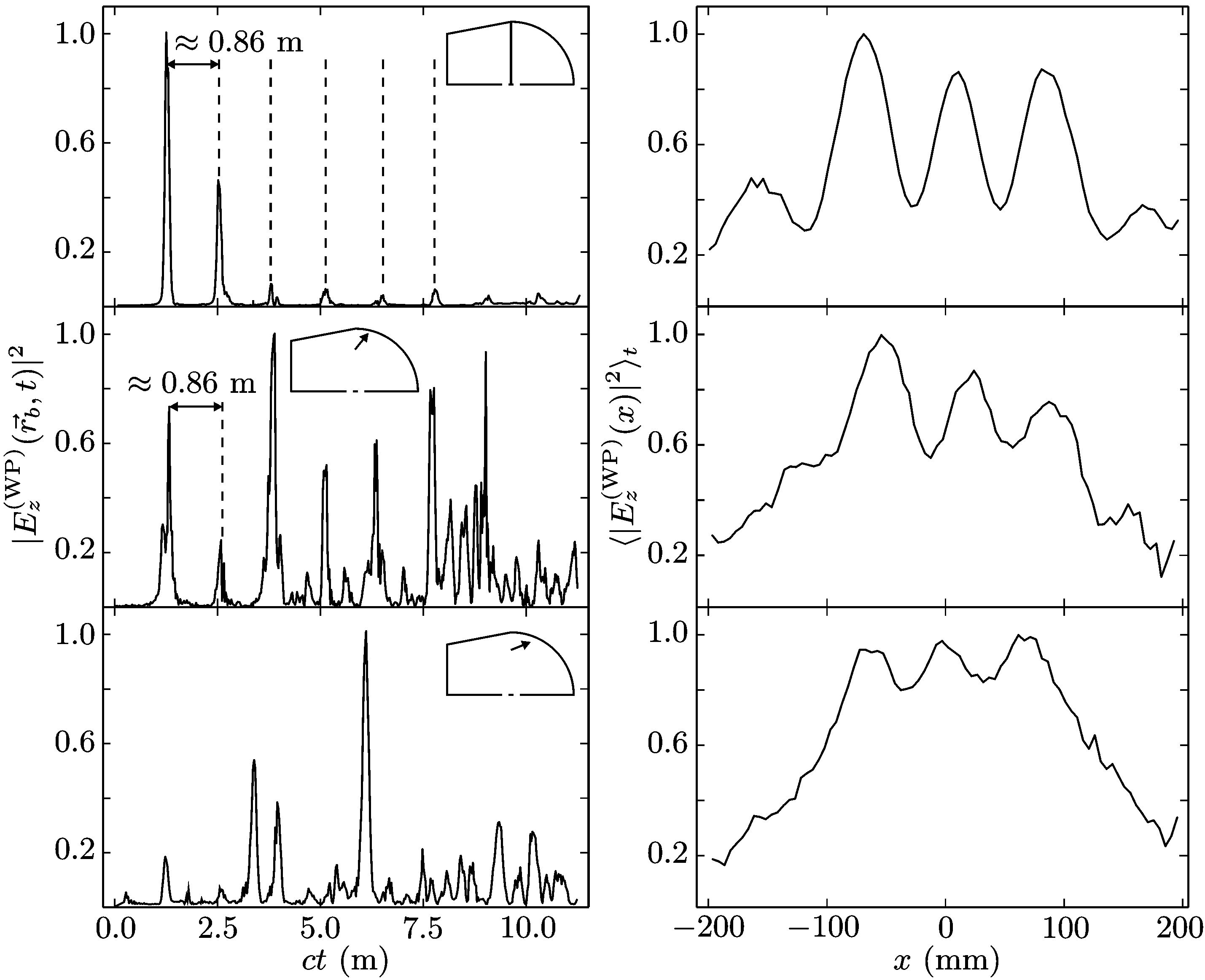} 
\caption{Time spectra (left panels) with antenna $b$ at $x = 0$ and time-averaged intensity (right panels) for a wave packet constructed according to \refeq{GauSta3} which starts with an initial angle of $270^\circ$ (top), $51^\circ$ (middle), and $22^\circ$ (bottom) in the stadium billiard. The dashed lines in the top left panel indicate the periodicity of the escape times. The parameters $\dist$, $\slit$, and $l$ are the same as in \reffig{fig:ScreenTime}.}
\label{fig:EchoesWpStad}
\end{center}
\end{figure*}

Three cases corresponding to different values of the initial angle $\alpha$ are analyzed in the following. Figure \ref{fig:EchoesWpRect} shows for the rectangular billiard in the left panels the time spectra and in the right panels the time-averaged intensity patterns for an initial wave packet with width $\sigma$ = 30 mm and \mbox{$\alpha$ = $270^\circ$} (top panels), $\alpha$ = $51^\circ$ (middle panels), and \mbox{$\alpha$ = $22^\circ$} (bottom panels) and antenna $b$  positioned at $x = 0$. The corresponding periodic orbits are shown in the insets. The parameters $\dist$, $\slit$, and $l$ are the same as in \reffig{fig:ScreenTime}. The time average was computed up to $75$~ns which corresponds to $ct = 22.5$~m. In the time spectra the positions of the peaks correspond to the time the initial wave pulse needs to reach the slits. For an initial angle $\alpha$ = $270^\circ$ the wave packet moves toward the slits along the vertical bouncing ball orbit, it travels back and forth reflecting at the upper and the lower edges of the billiard. The time spectrum exhibits a clear periodicity. The peaks appear in pairs since one part of the wave packet is sent to the back with an initial angle of $\alpha = 90^\circ$ and the other to the slits. The period of the escape times in the upper left panel agrees well with the length of the bouncing ball orbit, $0.95$ m. In the top right panel of \reffig{fig:EchoesWpRect} the time-averaged spatial intensity displays an interference pattern with visibility $(I_{\mathrm{max}}-I_{\mathrm{min}})/I_{\mathrm{max}}$ close to $100 \%$.

The initial angle $\alpha$ = $51^\circ$ corresponds to the periodic orbit shown in the inset of the left middle panel of \reffig{fig:EchoesWpRect}. The peak structure observed in the middle left panel does not show a clear periodicity. This is attributed to the spreading of the wave packet in additional directions (cf. \reffig{fig:freeProp}). The visibility of the corresponding interference pattern (see middle right panel of \reffig{fig:EchoesWpRect}) is \mbox{approximately $40 \%$}. The initial angle \mbox{$\alpha$ = $22^\circ$} corresponds to the orbit depicted in the inset of the bottom left panel of \reffig{fig:EchoesWpRect}. For this case the peaks appear stochastically. Still, the time-averaged spatial intensity depicted in the  bottom right panel shows interference patterns with visibility close to \mbox{$80 \%$}. The presence of a few prominent peaks, especially that at about $1.5$ m, in the time spectrum suggests that these dominate the interference pattern. However, the shape of the interference pattern is barely modified if the largest peak is removed. 

The same initial angles are chosen in the stadium billiard. The results are shown in \reffig{fig:EchoesWpStad}. The peaks in the  top left panel ($\alpha = 270^\circ$) again appear periodically, where the period is equal to the length of the vertical diffractive orbit shown in the inset, which is \mbox{$0.86$ m}. The visibility of the interferences is less than that for the rectangular billiard and is approximately \mbox{$60 \%$}. In addition, the peaks are slightly nonsymmetrical with respect to $x=0$. Thus, in spite of remains of interferences, there are visible differences to the rectangular billiard in the interference patterns (see the top right panel of \reffig{fig:EchoesWpRect}). The dominant peaks in the  middle left panel ($\alpha = 51^\circ$) display a periodic structure that again corresponds to the period of the vertical diffractive orbit, indicating that a part of the wave packet sticks in the vicinity of this periodic orbit. In addition, further narrow peaks in between are observed. The visibility of the interference pattern shown in the middle  right panel of \reffig{fig:EchoesWpStad} is moderately lower than the one for the rectangular billiard (see the middle right panel of \reffig{fig:EchoesWpRect}). In the  bottom left panel ($\alpha = 22^\circ$) the escape times exhibit no periodicity. The interferences disappear almost completely and the visibility is approximately $15 \%$. Generally, the differences between the intensity patterns originating from the rectangular and the stadium billiard with regular and chaotic dynamics, respectively, are especially large for initial angles close to $20^\circ$. A question that immediately arises is why in the regular case interferences are well pronounced only for certain initial angles.

\section{\label{conclusions}Conclusions}
We performed double-slit experiments with waves leaving a microwave billiard whose dynamics is either regular or chaotic. Microwaves were coupled into the resonators with differently arranged antennas, and the dependence of the interference patterns on a screen outside the billiard on its dynamics, the mode of excitation and the initial direction of the signal were analyzed. 

The emission of a wave pulse from a single antenna generates an omnidirectional pulse. Unless the system consisting of the billiard with the two slits and the emitting antenna is fully symmetric, the waves travel along different paths in the billiard before exiting through the slits. For an asymmetric setup the interference pattern measured on a screen at a certain distance from the lower edge of the billiard changes with time. Consequently, when averaging over time the interference patterns vanish. Thus, their emergence does not only depend on whether the dynamics inside the billiard is regular or chaotic, whereas the choice of a definite direction of the initial momentum of the wave packet seems to be crucial \cite{Fonte2006}. 

Furthermore, experiments with a square array configuration of \mbox{$5 \times 5$} emitting antenna positions were carried out. The superposition of the Fourier transforms of the spectra measured successively for each of these positions provides a method to create a directed wave packet. A test of it in free space indeed yielded propagation of the wave packet in a specific direction, although it spreads out quickly and has additional propagation directions. For the rectangular billiard with regular dynamics a clear interference pattern with $100\%$ visibility is only observed if the initial wave packet is sent along its symmetry line. In the stadium billiard with chaotic dynamics remnants of interferences are observed practically for all initial angles of the directed wave packet although the visibility is in general smaller than for the rectangular cavity. For angles close to $20^\circ$ the interferences are almost completely suppressed, whereas they appear with high visibility (about $80 \%$) for the rectangular one. The underlying mechanism for the formation of interferences is not fully understood since the high sensitivity of the patterns to the choice of the initial parameters of the wave packet does not allow to draw final conclusions. 

In conclusion, the numerical results presented by Casati and Prosen \cite{Casati2005} are only approximately reproduced for certain initial directions for both the rectangular and the stadium billiard. These discrepancies cannot be attributed to the different orders in time of the electromagnetic wave equation and the Schr{\"o}dinger equation since the interference patterns always result from a time average. It should be noted that in the simulations performed by Casati and Prosen the sharp distinction between patterns from billiards with regular and chaotic dynamics was only obtained for certain initial conditions, too \cite{Prosen2008}. This demonstrates that further investigations are necessary for an understanding of the interference patterns. Repeating the present measurements with visible light is desirable because a higher ratio between the cavity size and the wavelength can be realized, i.e., the system will be closer to the semiclassical limit. Preparations of appropriately shaped optical cavities with regular and chaotic dynamics, respectively, are underway.

\begin{acknowledgments}
We acknowledge T.\ Prosen and H.-J.\ St{\"o}ckmann for numerous suggestions and useful discussions. Special thanks go to R. Jakoby and his group from the Institute of Microwave Engineering at the TU Darmstadt for providing the anechoic chamber and helping with the measurements in it. One of us (P.\ O.\ I.) gratefully thanks the Deutscher Akademischer Austausch Dienst (DAAD) and the Fundaci\'{o}n La Caixa for financial support. This work was supported by the DFG through SFB634.
\end{acknowledgments}

\end{document}